# Asymmetric Cloaking Theory Based on Finsler Geometry

~ How to design true invisibility cloak with a scientific method ~

Tomohiro AMEMIYA [1], Daisuke NISHIYAMA [2] [†], and Masato TAKI [3]

[1] *Quantum Nanoelectronics Research Center, Tokyo Institute of Technology, 2-12-1 O-okayama, Meguro-ku, Tokyo 152-8552, Japan*
  E-mail: amemiya.t.ab@m.titech.ac.jp

[2] *Department of Physics, Tokyo Institute of Technology, 2-12-1 O-okayama, Meguro-ku, Tokyo 152-8551, Japan*

[3] *RIKEN Nishina Center for Accelerator-Based Science, 2-1 Hirosawa, Wako, Saitama 351-0198, Japan*
  E-mail: taki@riken.jp

**Abstract.** Is it possible to actually make Harry's invisibility cloaks? The most promising approach for realizing such magical cloaking in our real world would be to use transformation optics, where an empty space with a distorted geometry is imitated with a non-distorted space but filled with transformation medium having appropriate permittivity and permeability. An important requirement for true invisibility cloaks is *nonreciprocity*; that is, a person in the cloak should not be seen from the outside but should be able to see the outside. This invisibility cloak, or a nonreciprocal shield, cannot be created as far as we stay in conventional transformation optics. Conventional transformation optics is based on Riemann geometry with a metric tensor independent of direction, and therefore cannot be used to design the nonreciprocal shield. To overcome this problem, we propose an improved theory of transformation optics that is based on *Finsler geometry*, an extended version of Riemann geometry. Our theory shows that nonreciprocal shielding can be realized by covering cloaking space with transformation medium having anisotropic, nonreciprocal permittivity and permeability. This theory includes conventional Pendry's and Leonhardt's theories as special cases. We show the method for designing the spatial distribution of the permittivity and permeability required to make the nonreciprocal shield.

## 1. Introduction

Artificial materials such as plasmonic materials and metamaterials can provide extraordinary optical properties that cannot exist in nature [1-3]. This enables us to create novel optical devices that were previously considered impractical. One impressive example is an 'invisibility cloak,' a cloak that makes its wearer transparent and invisible to others. Magical invisibility cloaks, of course, belong to the world of fairy tales and science fictions. It is physically possible, however, to make a physical object seem invisible by directing incident light to avoid the object, flow around the object, and return undisturbed to its original trajectory.

One of favorable winds for this physical invisibility cloaks is the recent progress in the research of metamaterials. Metamaterials are artificial materials consisting of multiple structural elements, such as metal resonators, arranged periodically with a pitch smaller than the wavelength of light [4-7]. They can exhibit permittivity and permeability values that are not found in nature. Using metamaterials, it is possible to make a peculiar electromagnetic field surrounding an object we wish to hide, and therefore it would be possible to control the optical path around the object and make the object seem invisible.

The theoretical method of designing invisibility cloaks was first proposed by Pendry and coworkers in 2006 [8]. They proposed using an anisotropic medium to control light ray and calculated the spatial distribution of medium's permittivity and permeability that are needed for cloaking. At about the same time, Leonhardt made a proposal for invisibility cloaks that uses an isotropic medium [9, 10]. His proposal offers ease of actual device fabrication, and therefore would be promising for practical use, but has some essential problems because a mathematical theorem called as Nachman theorem shows that perfect invisibility cloaking without reflection and phase delay is impossible to realize with isotropic media [11, 12]. These two papers are purely theoretical papers that indicate the feasibility of invisibility cloaks, and they did not show the method of realizing actual materials with peculiar permittivity and permeability needed for fabricating invisibility cloaks. Since then, however, much effort has been expended in the development of metamaterials for invisibility cloaking, and many results of cloaking experiments using metamaterials have been reported [13-15].

---

[†] He recently founded Maple Systems Co., Ltd. which is a social application provider headquartered in Shinjuku, Tokyo, Japan.
  E-mail: d.nishiyama@maplesystems.co.jp

The point of invisibility cloaking is to make a closed region (or shielded region) that incident light circumvents without passing through nor reflecting on. An object placed in the shielded region appears to outside onlookers as if it does not exist. If a person hides himself within the shilded region, he will become invisible from the outside from all possible viewing angles. Howerver, a serious, critical problem occurs here. That is, no light enters the shielded region, and therefore the person hiding therein will not be able to see out.

To overcome this problem, we improve the theory of invisibility cloaks to incorporate directionality or unilaterality; that is, we propose a theory of asymmetric shielding (or asymmetric invisible cloaking). The asymmetric shield as referred to here means a device that has one-directional transparency, wherein a person inside the shielded region can see out, but people outside cannot see inside the shielded region. In other words, we will formulate a complete shielding theory, in which "they cannot see us, but we can see them completely."

Constructing an invisibility cloak requires the following four steps:
(1) First, suppose an appropriately distortrd virtual space, in which incident light is bent by spatial distortion to avoid and circumvent a given shielded region.
(2) Next, using transformation optics, imitate the distorted space with a non-distorted space (real space) that is filled with a medium having an appropriate spatial distribution of permittivity and permeability.
(3) Create a practical material structure that can realize the spatial distribution of permittivity and permeability determined above.
(4) Construct an actual invisibility cloak, using the proposed material structure.

We show the design theory of asymmetric shielding. More specifically, we will discuss steps 1 and 2 from the viewpoint of asymmetric shielding, and show a design methodology for the spatial distribution of permittivity and permeability. To realize asymmetric shielding, *directionality* is needed for permittivity and permeability; directionality means the properties including nonreciprocity in addition to anisotropy (i.e., directionality ⊃ anisotropy). Our theory is an expansion of existing invisibility cloaking and includes Pendry's and Leonhardt's theories as special cases. This paper is a purely theoretical paper, and practical methods of realizing a directional medium (the portion described in steps 3 and 4) will be set aside for some other occasion.

## 2. Designing asymmetric invisibility cloaks using Finsler geometry

At present, invisibility cloaks are designed as follows, with the aid of transformation optics. According to the theory of transformation optics, a distorted empty space (curvilinear vacuum space) is equivalent in terms of the propagation of light to a non-distorted space (flat space) filled with a transformation medium having an appropriate spatial distribution of permittivity and permeability. Therefore, to realize invisibility cloaking, a curvilinear virtual space that can make objects invisible is first designed, and then the spatial permittivity-permeability distribution in the medium is determined to make the flat real space equivalent to the curvilinear virtual space. In this transformation process, both spaces are described by Riemann geometry, and the permittivity-permeability distribution for the flat space is calculated from the Riemannian metric tensor of the curvilinear space. This method of transformation is very useful to design of conventional invisibility shields. This method, however, cannot be used for asymmetrical shielding we are considering. The reason is that the conventional theory of transformation optics is based on the Riemann geometry and has no concept of directionality. The metric of Riemann space is a function only of coordinates and does not depend on the differential of coordinates.

To deal with this problem, we propose introducing directionality to the theory of transformation optics. This can be performed using Finsler geometry [16] instead of Riemann geometry. Finsler geometry is a geometry where the distance between two adjacent points is a function of both the magnitude and direction of the vector connecting the two points. Finsler space is an expansion of Riemann space and is reduced to Riemann space if directionality is removed.

The fundamental quantity of Finsler space is the Finsler metric, $F(x,\mathbf{y})$, where $x$ represents a position vector in the space and $\mathbf{y}$ represents a direction vector at that position. It gives a geometrical length to a directed curve in the space; the infinitesimally small distance for $x \sim dx$ is given by $ds = F(x, dx)$. The Finsler metric is a constant for a non-directional, flat space (i.e., for a flat Riemann space). The metric tensor, $g_{ij}$, of the space is related with the Finsler metric as

$$F(x,\mathbf{y}) = \sqrt{g_{ij}(x,\mathbf{y}) y^i y^j} \tag{1}$$

or

$$g_{ij} = \frac{1}{2}\left(F^2\right)_{y^i y^j}. \tag{2}$$

Metric tensor $g_{ij}$ is a function of coordinates and their differentials [17, 18].

To develop transformation optics on Finsler spaces, we consider a curvilinear, vacuum Finsler space (Fig. 1(b)) and a non-directional flat space filled with a directional medium (Fig. 1 (c)). The directional medium means a medium whose permittivity and permeability are anisotropic and nonreciprocal.

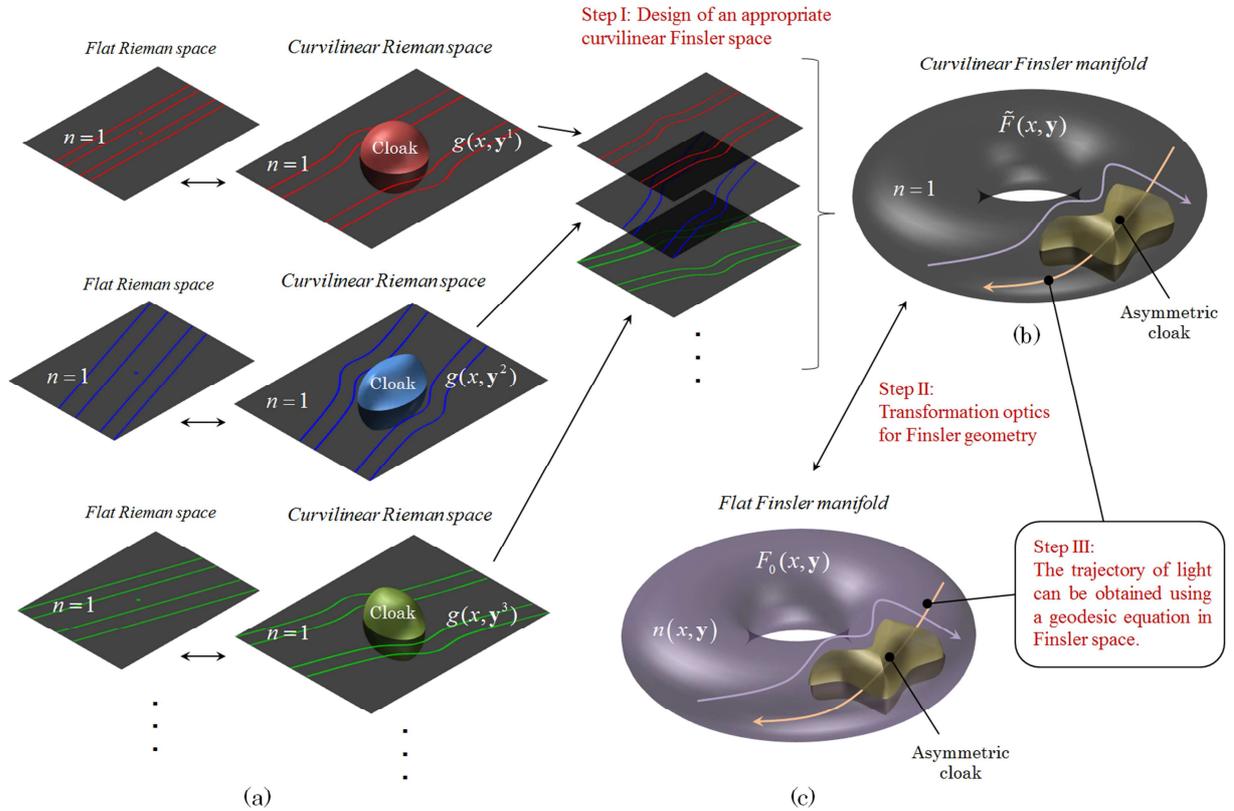

Fig. 1: Three kinds of manifolds (spaces) used in transformation optics based on Finsler geometry: (a) set of Rieman spaces; (b) curvilinear, vacuum Finsler space; (c) non-directional flat space filled with a directional medium. Closed curve in each figure indicate shielded regions.

(*Curvilinear vacuum Finsler space*)

Fig. 1(b) shows a curvilinear, vacuum Finsler space. If Fiesler metric $F(x,\mathbf{y})$ is set appropriately, the incident light circumvents the shielded region (indicated by a closed curve in the figure). A Finsler space is, when seen from one direction, equivalent to a curvilinear Riemann space with a uniform metric value. There is an infinite number of directions, and therefore a Finsler space is a set of an infinite number of Rieman spaces (Fig. 1 (a)).

(*Flat space with a directional medium*)

Fig. 1 (c) depicts a non-directional flat space filled with a directional medium. This space is also described by Finsler geometry, but it is equivalent to a flat Riemann space. To develop the theory of asymmetric shielding, we assume that the medium has a directional refractive index described by a tensor, $n(x,\mathbf{y})$, where $x$ and $\mathbf{y}$ represent position and direction. A flat space with refractive-index tensor $n(x,\mathbf{y})$ is equivalent, with regards to the propagation of light, to a curvilinear vacuum space whose line element $ds$ for a given set of $x$ and $\mathbf{y}$ is expressed by

$$ds = n(x,\mathbf{y})\|dx\| \tag{3}$$

If $n(x,\mathbf{y})$ is changed with an appropriate distribution in the flat space, light will be bent to circumvent the shielded region.

Using these two spaces, we can design an asymmetric shielding as follows.

(***Step 1***) First, deternime the shape of the curvilinear Finsler space (or determine Finsler metric $F(x,\mathbf{y})$) for an asymmetric shield. To do this, we consider a flat Riemann space for each direction and, by means of coordinate conversion, bend it to make a curvilinear Riemann space that acts as a desired shield in that direction (Fig. 1 (a)). Then, overlap the curvilinear Riemann spaces for possible directions to make a curvilinear Finsler space that has the desired property of asymmetric shielding (Fig. 1(b)).

(***Step 2***) Next, determine the spatial distribution of refractive index in the medium so that the flat space with the medium will become equivalent to the curvilinear Finsler space (Fig. 1 (c)). In other words, calculate medium's refractive index $n(x,\mathbf{y})$ that is needed for asymmetric shielding. From the refractive index and the non-reflection condition (described later), we can obtain the distribution profile of directional permittivity and permeability to realize asymmetric shielding.

(***Step 3***) Finally, confirm the opreation of the asymmetric shield by analyzing, with a geodesic equation, the trajectory of light in the flat space having the permittivity-permeability distribution we determined. This geodesic equation must be hold in Finsler space.

## 3. *Step 1*: Designing curvilinear Finsler space

Our first task is to decide the specific function of asymmetric shielding we need and to design an appropriate curvilinear Finsler space that produces the function. In other words, we first decide what kind of asymmetric shielding is needed, and then determine Finsler metric $F(x,\mathbf{y})$ so that the space will exhibit the property of the asymmetric shielding we need.

To do this, we propose in advance the method of giving the Finsler metric of space with directionality. As an simple example, we consider designing a three-dimensional shielding device shown in Fig. 2, where light takes a different path depending on its direction as follows.

(Fig. 2(R)) Light traveling leftward proceeds undisturbed. The person in the sielded region can therefore see the right-hand side. If the space were nondirectional, its metric tensor $g_{ij}^{R}$ would be the flat metric.

(Fig. 2(L1)) Light traveling rightward circumvents the shielded region. The person in the sielded region therefore seems invisible from the right-hand side. If the space were nondirectional, its metric tensor $g_{ij}^{L1}$ would be calculated with the conventional method using the algorithm wherein a space is distorted to expand a given point into a shielded region [7, 19, 20].

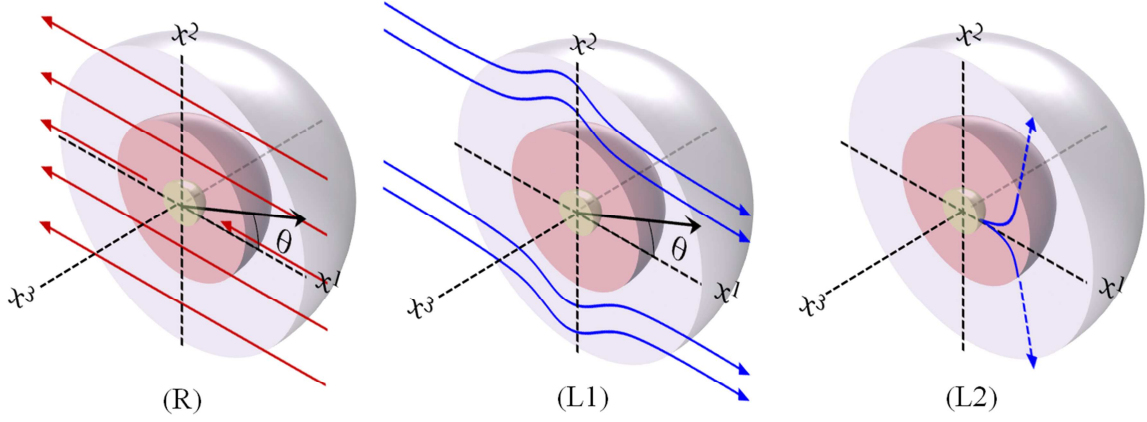

Fig. 2: Trajectory of light in and around a directional shielding device. In each figure, (1) the shielding device is enclosed by a large sphere, (2) the shielding region is enclosed by a medium sphere, and (3) a person in the shielding region is represented by a small sphere. (Left figure: R) Light teaveling leftward proceeds undisturbed. The person in the sielded region can therefore see the right-hand side; (Middle figure: L1) Light traveling rightward circumvents the shielded region. The person in the sielded region therefore seems invisible from the right-hand side; (Right figure: L2) Light from the person does not proceed rightward. This is needed to make the person invisible from the right side.

(Fig. 2(L2)) Light from the person does not proceed rightward. This is needed to make the person invisible from the right side. $g_{ij}^{L2}$ is the metric tensors for L2 in case that space were nondirectional. If the space were nondirectional, its metric tensor $g_{ij}^{L}$ would be calculated with the conventional method using the algorithm wherein a space is distorted to expand a given point into a shielded region.

The metric tensor of the directional space can be given as the convolution of $g_{ij}^{R}$, $g_{ij}^{L1}$, and $g_{ij}^{L2}$, and expressed by

$$g_{ij}(x,\theta) = f(\theta)g_{ij}^{L1} + (1-f(\theta))g_{ij}^{R} \quad \text{(Outside of device)} \tag{4}$$

$$g_{ij}(x,\theta) = f(\theta)g_{ij}^{L2} + (1-f(\theta))g_{ij}^{R} \quad \text{(Inside of device)}, \tag{5}$$

where $x$ is a position in the space, $\theta$ is a direction (angle) at the position, and $f(\theta)$ is a function that represents directionality. Metric $g_{ij}^{L2}$ can be obtained, for instance, with transformation $((r,\theta,\varphi) \to (r',\theta',\varphi'))$

$$r' = (r+r_0)\cosh\alpha(\theta), \quad \theta' = \theta, \quad \varphi' = \varphi \tag{6}$$

in a polar coordinate, where $r_0$ is a constant and $\alpha$ is a variable defined by

$$\tanh\alpha = \frac{2}{\pi}(\theta - \pi) \tag{7}$$

The asymmetric shielding can be achieved if we let, for instance,

$$f(\theta) = 0 \ (\pi/2 \le \theta < 3\pi/2), \ f(\theta) = 1 \ (0 \le \theta < \pi/2, \ 3\pi/2 \le \theta < 2\pi), \tag{8}$$

In practice, it would be better to use an appropriate smooth function for $f(\theta)$. From Eqs 1, 4 and 5, we obtain the Finsler metric $F(x,\mathbf{y})$ of the directional space as

$$F^2(x,\mathbf{y}) = \sum_{i,j=1}^{m} g_{ij}(x,\mathbf{y})y^i y^j = \sum_{i,j=1}^{m} \left(f(\theta)g_{ij}^{L} + (1-f(\theta))g_{ij}^{R}\right)y^i y^j. \tag{9}$$

On the basis of this idea, we can design the Finsler metric of a virtual curvilinear space for appropriate asymmetric shielding.

## 4. *Step II*: Calculating refractive-index distribution to imitate curvilinear Finsler space

Our next task is to replace the curvilinear Finsler space with the real flat space filled with a medium that has an appropriate spatial distribution of permittivity and permeability.

It is known that two conformally related Finsler manifolds, $(M, F)$ and $(M, \tilde{F})$, can be transformed into each other by equation

$$\tilde{F}(x, \mathbf{y}) = n(x) F(x, \mathbf{y}) \quad n \in C^{\infty}(M, R_+^*), \tag{10}$$

where $n(x)$ is a function of position $x$ [21]. In terms of physical optics, space $(M, \tilde{F})$ is equivalent to space $(M, F)$ with refractive-index tensor $n(x)$. To perform transformation between a curvilinear space and a flat space, let us expand Eq. 10 into

$$\tilde{F}(x, \mathbf{y}) = n(x, \mathbf{y}) F(x, \mathbf{y}), \tag{11}$$

where $n(x, \mathbf{y})$ is a function of position $x$ and direction $\mathbf{y}$ at the position. This equation means that, even if $(M, \tilde{F})$ is curvilinear and $(M, F)$ is flat, they can be transformed into each other provided that $n(x, \mathbf{y})$ is set to an appropriate function of position and direction. In other words, curvilinear space $(M, \tilde{F})$ can be replaced with flat space $(M, F)$ that is filled with a medium having an appropriate refractive-index tensor $n(x, \mathbf{y})$. To calculate $n(x, \mathbf{y})$ required for this purpose, let $\sim F(x, \mathbf{y})$ be the Finsler metric we determined in *Step I* and $F_0(x, \mathbf{y})$ be that of the real, flat space we live in. Because it is obvious that

$$\tilde{F}(x, \mathbf{y}) = \frac{\tilde{F}(x, \mathbf{y})}{F_0(x, \mathbf{y})} F_0(x, \mathbf{y}), \tag{12}$$

we obtain from Eqs. 11 and 1 that

$$n(x, \mathbf{y}) = \frac{\tilde{F}(x, \mathbf{y})}{F_0(x, \mathbf{y})} = \frac{\tilde{F}(x, \mathbf{y})}{\sqrt{g_{ij} y^i y^j}}, \tag{13}$$

where $g_{ij}$ is the metric tensor of the real, flat space and given by the unit matrix. The permittivity tensor $\varepsilon(x, \mathbf{y})$ and permittivity tensor $\mu(x, \mathbf{y})$ of the medium can be calculated from Eq. 13 and two equations

$$n(x, \mathbf{y}) = \sqrt{\varepsilon(x, \mathbf{y})} \sqrt{\mu(x, \mathbf{y})} \tag{14}$$

and

$$\frac{\sqrt{\mu(x, \mathbf{y})}}{\sqrt{\varepsilon(x, \mathbf{y})}} = C, \tag{15}$$

where $C$ is the surge impedance of space. Equation 15 shows a non-reflection condition to ensure that no reflection of incident light occurs at any point. If incident light enters from a vacuum to the shielding device, surge impedance $C$ has to be 1 (if enters from an air, to be approximately 1).

(*Consistency with existing theories*)

Before we move on to *Step III*, we would like to show that our theory includes existing theories as a special case where no directionality exists. For this purpose, we take an example of two-dimensional shielding devices that uses an anisotropic (but non-directional) medium according to Pendry and Leonhardt, and calculate the permittivity and permeability required for the medium. Our theory will give the same result as indicated by Pendry and Leonhardt.

Let us consider a curvilinear Riemann space for a non-directional invisibility cloak. The Finsler metric $\widetilde{F}(x,\mathbf{y})$ of the space is given by

$$\widetilde{F}(x,\mathbf{y}) = \sqrt{\widetilde{g}_{ij} y^i y^j}, \tag{16}$$

where $\widetilde{g}_{ij}$ is the metric tensor of the space and can be calculated using the existing algorithm wherein a point is expanded to form a shielded region [7, 19, 20]. For simplicity, we consider an example of a cylindrical transformation. Suppose the device transforms the radius $r'$ in electromagnetic space, but does not affect the cylindrical angle

$$r = r(r'), \quad \theta = \theta' \tag{17}$$

We calculate the metric tensor $\widetilde{g}_{ij}$ on the basis of $\{r', \theta'\}$-to-$\{r, \theta\}$ transformation. The result is

$$\widetilde{g}_{ij} = \begin{pmatrix} \frac{\partial r'}{\partial r} & \frac{\partial r'}{\partial \theta} \\ \frac{\partial \theta'}{\partial r} & \frac{\partial \theta'}{\partial \theta} \end{pmatrix} \begin{pmatrix} 1 & 0 \\ 0 & r'^2 \end{pmatrix} \begin{pmatrix} \frac{\partial r'}{\partial r} & \frac{\partial r'}{\partial \theta} \\ \frac{\partial \theta'}{\partial r} & \frac{\partial \theta'}{\partial \theta} \end{pmatrix}^t = \begin{pmatrix} \frac{\partial r'}{\partial r} & 0 \\ 0 & 1 \end{pmatrix} \begin{pmatrix} 1 & 0 \\ 0 & r'^2 \end{pmatrix} \begin{pmatrix} \frac{\partial r'}{\partial r} & 0 \\ 0 & 1 \end{pmatrix} = \begin{pmatrix} \left(\frac{\partial r'}{\partial r}\right)^2 & 0 \\ 0 & r'^2 \end{pmatrix}. \tag{18}$$

Refractive index tensor $n(x,\mathbf{y})$ can be calculated using Eqs. 13, 16, and 18, and given by

$$n(x,\mathbf{y}) = \frac{\widetilde{F}(x,\mathbf{y})}{\sqrt{g_{ij} y^i y^j}} = \frac{\sqrt{\left(\frac{\partial r'}{\partial r}\right)^2 (y^r)^2 + r'^2 (y^\theta)^2}}{\sqrt{(y^r)^2 + r^2 (y^\theta)^2}}. \tag{19}$$

The refractive index tensor under the non-reflection condition ($C=1$ in Eq. 15) is normally obtained using the permittivity and permeability of the medium, and given by

$$\mathrm{diag}(n_x^2, n_y^2, n_z^2) = \mathrm{diag}(\varepsilon_y \varepsilon_z, \varepsilon_z \varepsilon_x, \varepsilon_x \varepsilon_y). \tag{20}$$

This means that the refractive index in one direction of the local eigensystem of dielectric matrix does not depend on the ε and μ in that direction, but the dielectric properties in the orthogonal directions. This is because electromagnetic waves are transversal (their fields point orthogonally to the direction of propagation). Using Eqs. 19, 20, and the concept of the optical indicatrix, we can obtain the permittivity and permeability of the medium. If light enters the shielding device from a vacuum, the permittivity and permeability are given by

$$\varepsilon^i{}_j = \mu^i{}_j = diag\left(\frac{r'R}{r} \quad \frac{r}{r'R} \quad \frac{r'}{rR}\right). \tag{21}$$

This is equal to the result given by Pendry and Leonhardt. It is therefore clear that our theory includes their theory as a special case.

## 5. *Step III*: Confirming the trajectory of light

The design of the asymmetric shielding device was finished in Sect. 4. That is, we determined the spatial profile of directional permittivity and permeability for the medium to make an equivalent of the curvilinear Finsler space. However, the design was performed through the operation of metrics, and the trajectory of light does not appear explicitly. If we want to verify that light bends as designed in the space, we need to solve a geodesic equation in Finsler space.

For this purpose, let us derive the geodesic equation that holds in Finsler space. The length, $S$, of the trajectory, $X^i(t)$, in Finsler space is given by

$$S = \int_{t_1}^{t_2} F\left(X^i, \frac{dX^i}{dt}\right) dt. \qquad (22)$$

Here, $F$ satisfies Caratheodory condition

$$F(X, \lambda y) = \lambda F(X, y) \quad \text{for} \quad \lambda > 0. \qquad (23)$$

The geodesic equation can be derived using the variational method under this condition. If the trajectory changes minutely from $X^i(t)$ to $X^i(t) + \varepsilon v^i(t)$, then $S$ changes as

$$\begin{aligned} S(\varepsilon) &= \int_{t_1}^{t_2} F\left(X + \varepsilon v, \frac{dX}{dt} + \varepsilon \frac{dv}{dt}\right) dt = S(0) + \varepsilon \int_{t_1}^{t_2} \left(F_i v^i + F_{(i)} \frac{dv^i}{dt}\right) dt \\ &= S(0) + \left[F_{(i)} v^i\right]_{t_1}^{t_2} + \int_{t_1}^{t_2} \left(F_i - \frac{dF_{(i)}}{dt}\right) v^i dt \\ &= S(0) + \int_{t_1}^{t_2} \left(F_i - \frac{dF_{(i)}}{dt}\right) v^i dt \end{aligned} \qquad (24)$$

where

$$F_{(i)} \equiv \frac{\partial F(x, \mathbf{y})}{\partial y^i}. \qquad (25)$$

Therefore,

$$\left.\frac{dS}{d\varepsilon}\right|_{\varepsilon=0} = \int_{t_1}^{t_2} \left(F_i - \frac{dF_{(i)}}{dt}\right) v^i dt. \qquad (26)$$

In order for $S(\varepsilon)$ to become minimum at $\varepsilon = 0$, it must be that $dS(\varepsilon)/d\varepsilon = 0$. Because this has to hold for arbitrary $v^i$, it is that

$$F_i - \frac{dF_{(i)}}{dt} = 0. \qquad (27)$$

Using $\varepsilon \equiv F^2/2$, we rewrite Eq. 27 as

$$\varepsilon_i - \frac{d\varepsilon_{(i)}}{dt} = FF_i - \frac{d}{dt}(FF_{(i)}) = -F_{(i)} \frac{dF}{dt}. \qquad (28)$$

Using metric tensor $g_{ij}$ of the curvilinear Finsler space, the left side of Eq. 28 can be rewritten a

$$\begin{aligned} \varepsilon_i - \frac{d\varepsilon_{(i)}}{dt} &= \frac{1}{2} \frac{\partial g_{jk}}{\partial X^i} y^j y^k - \left(\frac{\partial g_{ji}}{\partial X^k} \frac{dX^k}{dt} + \frac{\partial g_{ji}}{\partial y^k} \frac{dy^k}{dS}\right) y^j - g_{ji} \frac{d^2 X^j}{dt^2} \\ &= -g_{ji} \frac{d^2 X^j}{dt^2} + \frac{1}{2}\left(\frac{\partial g_{jk}}{\partial X^i} - \cdots\right) y^j y^k \\ &= -g_{ji} \cdot (\text{equation for ordinary land survey line}) \end{aligned} \qquad (29)$$

The right side of Eq. 28 can be rewritten using $g_{ij}$ as

$$-F_{(i)} \frac{dF}{dt} = -\frac{\partial}{\partial y^i}\left(\sqrt{2\varepsilon}\right) \frac{d\left(\sqrt{2\varepsilon}\right)}{dt} = -\left(\sqrt{2}\right)^2 \frac{1}{2\sqrt{\varepsilon}} \frac{\partial \varepsilon}{\partial y^i} \frac{1}{2\sqrt{\varepsilon}} \frac{d\varepsilon}{dt} = -\frac{1}{2\varepsilon} g_{ji} y^j \frac{d\varepsilon}{dt}. \qquad (30)$$

Using Eqs. 29-31, we can obtain the geodesic equation for Finsler space. It is given by

$$\frac{d^2 X^i}{dt^2} + \Gamma^i{}_{jk} \frac{dX^j}{dt}\frac{dX^k}{dt} = \frac{d \log F(x,\mathbf{y})}{dt}\frac{dX^i}{dt}. \qquad (31)$$

If the medium has no directionality, Eq. 31 is reduced to the geodesic equation of Riemann space. The trajectory of light can be calculated by substituting the Finsler metric $F(x, \mathbf{y})$, obtained in *Step I*, into Eq. 31. It will be explicitly shown that incident light avoids the shielding region, flows around the shielding region, and returns undisturbed to its original trajectory.

## 6. Summary

We proposed the design theory of asymmetric invisibility shielding. Asymmetric shielding needs the concept of directionality, namely, a combination of anisotropy and nonreciprocity. To deal with this directionality, we expanded existing transformation optics based on Riemann geometry and formulated a new theory based on Finsler geometry. On the basis of this theory, we showed, in Sect. 3, the method of designing a virtual, curvilinear directional space for asymmetric shielding, and then determined, in Sect. 4, the spatial profile of directional permittivity and permeability for the medium in the real, flat space to make an equivalent of the curvilinear space. We derived, in Sect. 5, the geodesic equation in Finsler space to confirm the trajectory of light around the asymmetric shield. Our theory includes Pendry's and Leonhardt's theories as a special case that has no directionality. Our theory would enable to design invisibility cloaks that appear in Harry Potter's world, though omnidirectionally asymmetric invisibility cloaks are theoretically impossible.

An open question is whether it is possible to create a medium that has the property of nonreciprocal permittivity and permeability. The nonreciprocity can probably be realized to some degree, at microwave frequencies, using metamaterials consisting of integrated Ferrite [22]. At frequencies of light, however, it seems quite difficult at present to obtain nonreciprocal material that can be used for asymmetric shielding. Can we really create such profound material in the near future? Part of the secret might be in the magic of Hogwarts School of Witchcraft and Wizardry.